\documentclass[showpacs,aps,pre]{revtex4}
\usepackage{graphicx}
[1999/03/29 PSNFSS v.7.2
Times font as default roman
: S Rahtz]

\begin{document}
\renewcommand{\thefootnote}{\fnsymbol{footnote}}
\sloppy
\newcommand{\rp}{\right)}
\newcommand{\lp}{\left(}
\newcommand \be  {\begin{equation}}
\newcommand \ba {\begin{eqnarray}}
\newcommand \ee  {\end{equation}}
\newcommand \ea {\end{eqnarray}}

\title{Optimal Prediction of Time-to-Failure from Information 
Revealed by Damage}
\thispagestyle{empty}

\author{D. Sornette$^{1,2}$ and J.V. Andersen$^{1,3}$}
\affiliation{$^1$ Laboratoire de Physique de la Mati\`ere Condens\'ee,
CNRS UMR 6622 and Universit\'e de Nice-Sophia Antipolis, 06108
Nice Cedex 2, France}
\affiliation{$^2$ Institute of Geophysics and Planetary Physics
and Department of Earth and Space Sciences,
University of California, Los Angeles, CA 90095}
\affiliation{$^3$ U.F.R. de Sciences Economiques, Gestion, Math\'ematiques
et Informatique, CNRS UMR 7536 and Universit\'e Paris X-Nanterre, 92001
Nanterre Cedex}

\email{vitting@unice.fr, sornette@moho.ess.ucla.edu}

\date{\today}

\begin{abstract}
We present a general prediction scheme of failure times based on
updating continuously with time the probability for failure of the
global system, conditioned on the information revealed on the
pre-existing idiosyncratic realization of the system by the damage that
has occurred until the present time. Its implementation on a simple
prototype system of interacting elements with unknown random lifetimes
undergoing irreversible damage until a global rupture occurs shows that
the most probable predicted failure time (mode) may evolve
non-monotonically with time as information is incorporated in the
prediction scheme. In addition, both the mode, its standard deviation
and, in fact, the full distribution of predicted failure times exhibit
sensitive dependence on the realization of the system, similarly to 
``chaos'' in spinglasses, providing a
multi-dimensional dynamical explanation for the broad distribution of
failure times observed in many empirical situations.
\end{abstract}


\maketitle

Systems of connected and interacting elements often fail through a
self-organizing cascade process. Predicting the remaining lifetime of a
complex structure or the precise time of failure remains an unsolved
problem for all applications (engineering structures,
materials, earthquakes, grids, networks, groups and so on),
notwithstanding its huge importance and overwhelming consequences.
Different strategies include deterministic modeling, stochastic one body
or many body approaches, computational intelligence methods, and many
other classifiers and pattern recognition techniques, all with
limitations and lack of sufficient understanding of the underlying
physical mechanisms. A major problem is that failure of a given system
is highly history- and sample-dependent: in contrast with
standard statistical physics, the problem is not to calculate an
ensemble averaged thermodynamic property but to obtain a precise statement
for each single idiosyncratic realization. This difficulty is bypassed
for instance in the strategy which consists in viewing material 
rupture as a kind of
universal critical transition \cite{guarinocreep} (see however \cite{Aetal95}),
which is based on the
hope, which is partially supported by experiments, that a large 
system may behave
like a typical realization with a kind of self-averaging property. 
But this misses
the real practical challenge which is to detect the possible 
existence of flaws,
either pre-existing or self-organized which are
known to create a great variability of the lifetimes from one sample 
to the next.
In addition, existing methods are often mute on the limits of predictability
and on the sensitivity to various elements of the system under consideration.

Here, we analyze this prediction problem with a simple prototype of
interacting elements with unknown random lifetimes undergoing
irreversible damage until a global rupture occurs,
the so-called time-dependent hierarchical fiber bundle model \cite{Newmandyn}. By
obtaining the absolute best prediction scheme in a probabilistic
sense, we are able to cast new
light on the above questions. Consistent with
the information usually available in realistic situations,
we assume the knowledge of only the statistical properties of the constituting
elements but not of their specific realizations. We use the physics of
their interaction to develop the prediction scheme. The key idea is
to update continuously with time the conditional probability for failure
of the global system, conditioned on the information revealed by the
damage that has occurred until the present time. Continuously
collecting information
on the on-going damage progressively reveals key information on the 
pre-existing
idiosyncratic realization of the system which can be gradually
integrated in a better and better probabilistic prediction.

Consider a hierarchical structure of elements with $N$ levels loaded
with a stress $\sigma$ per element. The
first level is made of the individual elements,
the second level is made of pairs of elements, the third level is 
made of pairs of pairs
and so on. 
This defines a discrete hierarchical tree of local
coordination $2$ (the results below are easy to extend to any coordination).
This topology impacts the dynamics of failure in the following way.
When one of the two bundles
of a given pair fails, its stress load is transfered instantaneously 
to the surviving
bundle, such that its load is doubled. When this bundle breaks, its
load is transfered to the pair of bundles associated to it if this 
second pair is still present.
Otherwise, it is transfered to the pair of two pairs linked at the 
next hierarchical
level. Given some
stress history $\sigma(t'), t'\geq0$, an element is assumed to break 
at some fixed random time,
where the probability that this random time takes a specific value $t$
is specified by its cumulative distribution function
\be
F_0(t)\equiv\int_0^t P_0(t')dt'=1-\exp\left\{-\kappa\int_0^t
[\sigma(t')]^{\rho} {\rm d}t'\right\}~.
\label{pdfggh}
\ee
This amounts to considering an element failure as a conditional Poisson process
with an intensity which is function of all the past stress history
weighted by the stress amplification exponent $\rho>0$.
Applied to material failure, this
law captures the physics of failure due to stress corrosion, to
stress-assisted thermal activation and to damage.
A system of $2^N$ elements is fully specified by attributing to each
element $i=1, ..., 2^N$ at the beginning of their history a fixed failure
time $t_i$ taken from the distribution (\ref{pdfggh}). The failure time $t_i$
is by definition the time at which the element $i$ would have broken
if the stress had stayed constant equal to the initial value $\sigma$.
But, the elements are coupled through the hierarchical load transfer rule
defined above. As a consequence of the
hierarchical structure of the load transfers occuring at each 
rupture, the stress applied
to a given element may increase, leading to a shortening of its lifetime.
Consider a pair of bundles with lifetimes $t_1<t_2$. At time $t=t_1$, 
when the first
bundle breaks down, its load is transfered to the second bundle. It is
easy to show from (\ref{pdfggh}) that this leads to a reduction of
its lifetime to \cite{Newmandyn}
\be
t_{12}=t_1+\alpha(t_2-t_1) <t_2~, ~~~~\alpha = 2^{-\rho}~.
\label{alphaa}
\ee
This law applies for any realization of lifetimes at all levels 
within the hierarchy
and forms the basis for our derivation below.

In order to mimic a real-life situation, we consider a creep experiment
of our hierarchical system, such that at time $0$, a stress $\sigma$
is applied. We have no access to the specific individual lifetimes
of the individual constituting elements, only to their probability
density function (PDF) $P_0(x)$, as in
a real experiment. At time passes, damage occurs, that is, elements break,
thus revealing their initial lifetimes or combination thereof. The 
situation becomes
rapidly complicated because of the interactions between the elements
through the hierarchical stress-load redistribution, as the damage 
spreads across the levels
of the hierarchy. In a real-life experiment, the damage in a material sample is
monitored for instance by acoustic emissions, with both time and
space localization.
In order to construct our prediction scheme, we just need to
construct the prediction scheme for a system of
four elements (or bundles) with a priori unknown initial lifetimes
$t_1, t_2, t_3$ and $t_4$, whose PDFs are known.
In the case where
each bundle reduces to an element of level 1,
the PDF's are identical and equal to $P_0(x)$, as we assume
that the elements have i.i.d. lifetimes. However, for the case of
four bundles of arbitrary level $j>0$, the PDF's of their lifetime
are a priori distinct and result
from the PDFs of the elementary elements at the first level
combined with the specific history of the damage until time $t$ undergone
by each bundle, as we now explain.

{\bf Prediction in absence of revealed damage}.
Let $P_i(t_i)$ denote the PDF of the lifetimes of element (or bundle)
$i$, with $i=1, .., 4$. If we knew
$t_1$ and $t_2$, we would determine the lifetime of the pair as
$t_{(1,2)} = {\rm Min}[t_1, t_2] + \alpha \left( {\rm Max}[t_1, t_2] 
- {\rm Min}[t_1, t_2]\right)$,
according to (\ref{alphaa}) (and similarly for the pair $(3,4)$). But
$t_1$ and $t_2$ are unknown, and the
best we can do is to calculate the PDF of $t_{(1,2)}$ at some given time $t$.
Conditioned on the fact that no element has failed, we have
\be
P_{(1,2),t}\left(t_{(1,2)}\right) =
  {1 \over \alpha} \int_t^{t_{(1,2)}} dt_1 ~{\tilde P}_{1,t}(t_1)
{\tilde P}_{2,t_1} \left( \left[t_{(1,2)}- (1-\alpha) t_1\right]/\alpha \right)
+ (1 \leftrightarrow 2)~,
\label{nmgjkaaaals}
\ee
where
${\tilde P}_{i,t}(t_i) = {P_i(t_i) \over \int_t^{\infty} P_i(x) dx}$
is the conditional PDF's of element $i$, given that it
has not yet broken at time $t$. The second contribution $(1 \leftrightarrow 2)$
in (\ref{nmgjkaaaals})
corresponding to $t_1>t_2$ is obtained
from the first contribution corresponding to $t_1<t_2$
by exchanging the two indices $1$ and $2$. We check that
$P_{(1,2),t}\left(t_{(1,2)}\right)$
is normalized to unity over the time interval from $t$ to $\infty$
by using the identity
$\int_t^{\infty} dt_{(1,2)} \int_t^{t_{(1,2)}} dt_1 = \int_t^{\infty} dt_1
\int_{t_1}^{\infty} dt_{(1,2)}$ and the change of variable
$t_{(1,2)} \to u \equiv \left[t_{(1,2)}- (1-\alpha) t_1\right]/\alpha$.
The PDF $P_{(1,2,3,4),t}\left(t_{(1,2,3,4)}\right)$ of the lifetimes 
$t_{(1,2,3,4)}$ of the
group of four elements at time $t$ conditioned on the fact that no element
has ruptured until $t$ has the same structure as
(\ref{nmgjkaaaals}) with the substitutions $1 \to (1,2)$ and $2 \to (3,4)$.

{\bf Using the knowledge that one element failed at time $t^*$}.
Suppose we record the failure of the element $1$ at time $t^*$, i.e., its
initially unknown lifetime $t_1$
is suddenly revealed: $t_1=t^*$. Conditioned on this information revealed
at time $t^*$, we know proceed to derive how this impacts the prediction
of the lifetime of the four elements, changing 
$P_{(1,2,3,4),t}\left(t_{(1,2,3,4)}\right)$
into a conditional PDF $P_{(1,2,3,4),t^*}\left(t_{(1,2,3,4)}\right)$.
Indeed, the failure of element $1$ at $t^*$ immediately changes
$P_{(1,2),t}\left(t_{(1,2)}\right)$ for the rupture time
of the pair $(1,2)$ (i.e., of element 2 given that element 1 has broken) from
expression (\ref{nmgjkaaaals}) to
\be
P_{(1,2),t^*}\left(t_{(1,2)}^*\right) = {1 \over \alpha} {\tilde 
P}_{2,t^*}\left(\left[t_{(1,2)}^*
-(1-\alpha) t^*\right]/\alpha\right)~.
\label{mfwdl}
\ee
Expression (\ref{mfwdl}) derives from
(\ref{nmgjkaaaals}) by replacing ${\tilde P}_{1,t}(t_1)$ by $\delta(t_1-t^*)$
to express the certain knowledge of the failure time of element $1$. It can
also be interpreted as the change of the failure time of element 2
from $t_2$ to $t^* +\alpha(t_2-t^*)$ by the stress transfer from element 1 to
element 2 occurring at $t^*$ (and with the proper normalization of the
distribution). The gain in prediction accurary described below
is due to the fact that the variance of $P_{(1,2),t^*}\left(t_{(1,2)}^*\right)$
given by (\ref{mfwdl}) is smaller than that of 
$P_{(1,2),t}\left(t_{(1,2)}\right)$
given by (\ref{nmgjkaaaals}).

In contrast with the previous case leading to $P_{(1,2,3,4),t}\left(t_{(1,2,3,4)}\right)$
when no failure has occurred yet,
the two pairs $(1,2)$ and $(3,4)$ do not play a symmetric role
and two scenarios can occur for times greater than
$t^*$, given that element $1$ has broken at $t^*$.
Scenario 1 is that element $2$ fails first, followed by the rupture 
of second pair $(3,4)$.
This scenario contains both the case where element $2$ breaks first
and then $(3,4)$ and the case when element $3$ (or $4$) breaks first, 
then element $2$ fails
and then element $4$. The probability of this scenario is
\ba
{\rm Pr}[t_{(1,2)}^* < t_{(3,4)}^*] &=& \int_{t^*}^{\infty} d t_{(1,2)}^*~
P_{(1,2),t^*}\left(t_{(1,2)}^*\right) \int_{t_{(1,2)}^*}^{\infty} d 
t_{(3,4)}^*~
P_{(3,4),t^*}\left(t_{(3,4)}^*\right)~.
\label{mvmala}
\ea
In the final
calculation of ${\rm Pr}[t_{(1,2)}^* < t_{(3,4)}^*]$, we must use
the fact that $P_{(1,2),t^*}\left(t_{(1,2)}^*\right)$ is given by 
(\ref{mfwdl}).
Scenario 2 is that the second pair $(3,4)$ breaks first, followed
by the failure of element $2$. This occurs with a probability
${\rm Pr}[t_{(1,2)}^* > t_{(3,4)}^*] =
1- {\rm Pr}[t_{(1,2)}^* < t_{(3,4)}^*]$

Conditioned on the fact that the rupture follows the first
scenerio ($t_{(1,2)}^* < t_{(3,4)}^*$),
the PDF for the failure time
$t_{(1,2,3,4)}^*$ of the whole four-element system is
\be
P_{(1,2,3,4),t^*}^{{\rm sc}~1}\left(t_{(1,2,3,4)}^*\right) =
{1 \over \alpha} \int_{t^*}^{t_{(1,2,3,4)}^*} dt_{(1,2)}^* ~
P_{(1,2),t^*}\left(t_{(1,2)}^*\right) 
P_{(3,4),t^*}\left(\left[t_{(1,2,3,4)} -(1-\alpha)
t_{(1,2)}^*\right]/\alpha \right)~.
\label{mgjsl}
\ee
Here, the PDF for the failure time of the first
pair $(1,2)$ is changed into $P_{(1,2),t^*}\left(t_{(1,2)}^*\right)$ given by
(\ref{mfwdl}).  We can thus rewrite (\ref{mgjsl}) as
\be
P_{(1,2,3,4),t^*}^{{\rm sc}~1}\left(t_{(1,2,3,4)}^*\right) =  {1 
\over \alpha^2}
\int_{t^*}^{t_{(1,2,3,4)}^*}
dt_{(1,2)}^* ~ {\tilde P}_{2,t^*}\left(\left[t_{(1,2)}^*
-(1-\alpha) t^*\right]/\alpha\right) ~ 
P_{(3,4),t^*}\left(\left[t_{(1,2,3,4)} -(1-\alpha)
t_{(1,2)}^*\right]/\alpha \right)~.
\label{mgjaaasl}
\ee

Conditioned on the fact that the rupture follows the second
scenerio ($t_{(1,2)}^* > t_{(3,4)}^*$)
the PDF for the failure time
$t_{(1,2,3,4)}^*$ of the whole four-element system is
\be
P_{(1,2,3,4),t^*}^{{\rm sc}~2}\left(t_{(1,2,3,4)}^*\right) =
{1 \over \alpha^2} \int_{t^*}^{t_{(1,2,3,4)}^*} dt_{(3,4)}^* ~
P_{(3,4),t^*}\left(t_{(3,4)}^*\right)
{\tilde P}_{2,t^*}\left( {t_{(1,2,3,4)}^* \over \alpha^2} -
{(1-\alpha) \over \alpha^2} t_{(3,4)}^* - {1-\alpha \over \alpha} t^* \right)~,
\label{mvjkssss}
\ee
where $P_{(1,2),t^*}\left(t_{(1,2)}^*\right)$ is given by
(\ref{mfwdl}).

Combining both scenarios yields the
PDF for the failure time
$t_{(1,2,3,4)}^*$ of the four-element system:
\be
P_{(1,2,3,4),t^*}\left(t_{(1,2,3,4)}^*\right) =
  P_{(1,2,3,4),t^*}^{{\rm sc}~1}\left(t_{(1,2,3,4)}^*\right)
+  P_{(1,2,3,4),t^*}^{{\rm sc}~2}\left(t_{(1,2,3,4)}^*\right)~.
\label{mfmeless}
\ee
where the two terms in the r.h.s. of (\ref{mfmeless}) are given respectively by
(\ref{mgjaaasl}) and (\ref{mvjkssss}). We verify that
the PDF $P_{(1,2,3,4),t^*}\left(t_{(1,2,3,4)}^*\right)$
is normalized to unity as
$\int_{t^*} P_{(1,2,3,4),t^*}\left(t_{(1,2,3,4)}^*\right) d t_{(1,2,3,4)}^* =
{\rm Pr}[t_{(1,2)}^* < t_{(3,4)}^*] + {\rm Pr}[t_{(1,2)}^* > t_{(3,4)}^*] =1$,
since the integral of (\ref{mgjaaasl}) gives (\ref{mvmala}),
and the integral of (\ref{mvjkssss}) gives the complement to $1$, using
the identity $\int_{t^*}^{+\infty} dy \int_{t^*}^y dx =
\int_{t^*}^{+\infty} dx \int_{x}^{+\infty} dy$ and a change of variable.

{\bf Two elements are broken in the same pair (i.e. scenario 1 is fulfilled)}.
Suppose that element 2 breaks at some later time $t^{\dagger} > t^*$
before the rupture of the pair $(3,4)$. This rupture reveals
a new information which can be exploited to improve the prediction of the
rupture time of the 4-element bundle. Indeed, expression (\ref{mfmeless}) is
changed into
\be
P_{(1,2,3,4),t^*,t^{\dagger}}\left(t_{(1,2,3,4)}^{*,\dagger}\right) =
{P_{(3,4),t^{\dagger}}\left(\left[t_{(1,2,3,4)}^{*,\dagger} -(1-\alpha)
t^{\dagger}\right]/\alpha \right) \over
\int_{t^{\dagger}}^{\infty} dx~P_{(3,4),t^{\dagger}}\left(\left[x -(1-\alpha)
t^{\dagger}\right]/\alpha \right)}  ~,~~~~{\rm for}~~ t \geq t^{\dagger}~.
\label{mgmle}
\ee
This corresponds to a considerable decrease of uncertainty: first,
scenario 2 is now excluded and, second, the distribution (\ref{mgjsl})
is collapsed similarly to the process leading to (\ref{mfwdl}) at time $t^*$.
The denominator ensures the normalization of
$P_{(1,2,3,4),t^*,t^{\dagger}}\left(t_{(1,2,3,4)}^{*,\dagger}\right)$
over the interval $[t^{\dagger}, +\infty]$ and expresses the
fact that $P_{(1,2,3,4),t^*,t^{\dagger}}\left(t_{(1,2,3,4)}^{*,\dagger}\right)$
is a distribution of failure times conditioned to the failure time being
larger than $t^{\dagger}$.
The PDF $P_{(3,4),t^{\dagger}}$ contains the information of whether
element 3 or 4 (but not both) have ruptured in the mean time, according
to a derivation similar to that leading to (\ref{mfwdl}) after the rupture
of element 1.

{\bf Two elements are broken, one in each of the pairs (1,2) and (3,4)}.
In this case the prediction of the rupture time is given by expression
(\ref{mfmeless}) but the knowledge that a element broke in $(3,4)$ means
that $P_{(3,4)}$ has to be replaced by the expression (\ref{mfwdl}),
with a change of indices $(1,2) \rightarrow (3,4)$ in (\ref{mfwdl}).

{\bf Three elements are broken with scenario 2}.
Suppose that the pair $(3,4)$ breaks at some time $t^{\dagger} > t^*$
before the failure of element 2. Then, again the prediction of the rupture
time of the 4-element bundle is improved according to
\be
P_{(1,2,3,4),t^*,t^{\dagger}}\left(t_{(1,2,3,4)}^{*,\dagger}\right) =
{ {\tilde P}_{2,t^*,t^{\dagger}}\left( {t_{(1,2,3,4)}^{*,\dagger} 
\over \alpha^2} -
{(1-\alpha) \over \alpha^2} t^{\dagger} - {1-\alpha \over \alpha} t^* 
\right) \over
\int_{t^{\dagger}}^{\infty} dx~
{\tilde P}_{2,t^*,t^{\dagger}}\left( {x \over \alpha^2} -
{(1-\alpha) \over \alpha^2} t^{\dagger} - {1-\alpha \over \alpha} t^* \right) }
~,~~~~{\rm for}~~ t \geq t^{\dagger}~.
\label{mgmhmrl}
\ee
The denominator ensures the normalization of
$P_{(1,2,3,4),t^*,t^{\dagger}}\left(t_{(1,2,3,4)}^{*,\dagger}\right)$
over the interval $[t^{\dagger}, +\infty]$ and expresses the
fact that $P_{(1,2,3,4),t^*,t^{\dagger}}\left(t_{(1,2,3,4)}^{*,\dagger}\right)$
is a distribution of failure times conditioned to the failure time being
larger than $t^{\dagger}$.

{\bf Three elements are broken with scenario 1}
The prediction of the rupture time is then given by
expression (\ref{mgmle}) but the knowledge that a element broke in $(3,4)$
means that
$P_{(3,4)}$ has to be replaced by the expression (\ref{mfwdl}),
with a change of indices $(1,2) \rightarrow (3,4)$ in (\ref{mfwdl}).

It is straightforward to iterate this enumeration for a system of arbitrary
size $2^N$. Here, we present results obtained for a system of $16$ elements,
with identical exponential distributions of lifetimes. In order to calculate
the PDF of the lifetime $t_c$ of the whole system, we decompose
it into four bundles of $4$ elements each, for which we calculate their
corresponding PDFs. The four PDFs for
each 4-bundle in turn take the role of the $\tilde{P_{i,t}}$ used
in the previous calculations of the PDF for the
total bundle of four 4-bundles. It is important to stress
that, even though the lifetimes of the individual
elements are i.i.d., the PDFs of the four 4-bundles remain the same
only as long as no individual element has broken and then diverge
as damage grows.

\begin{figure}
\includegraphics[width=14cm]{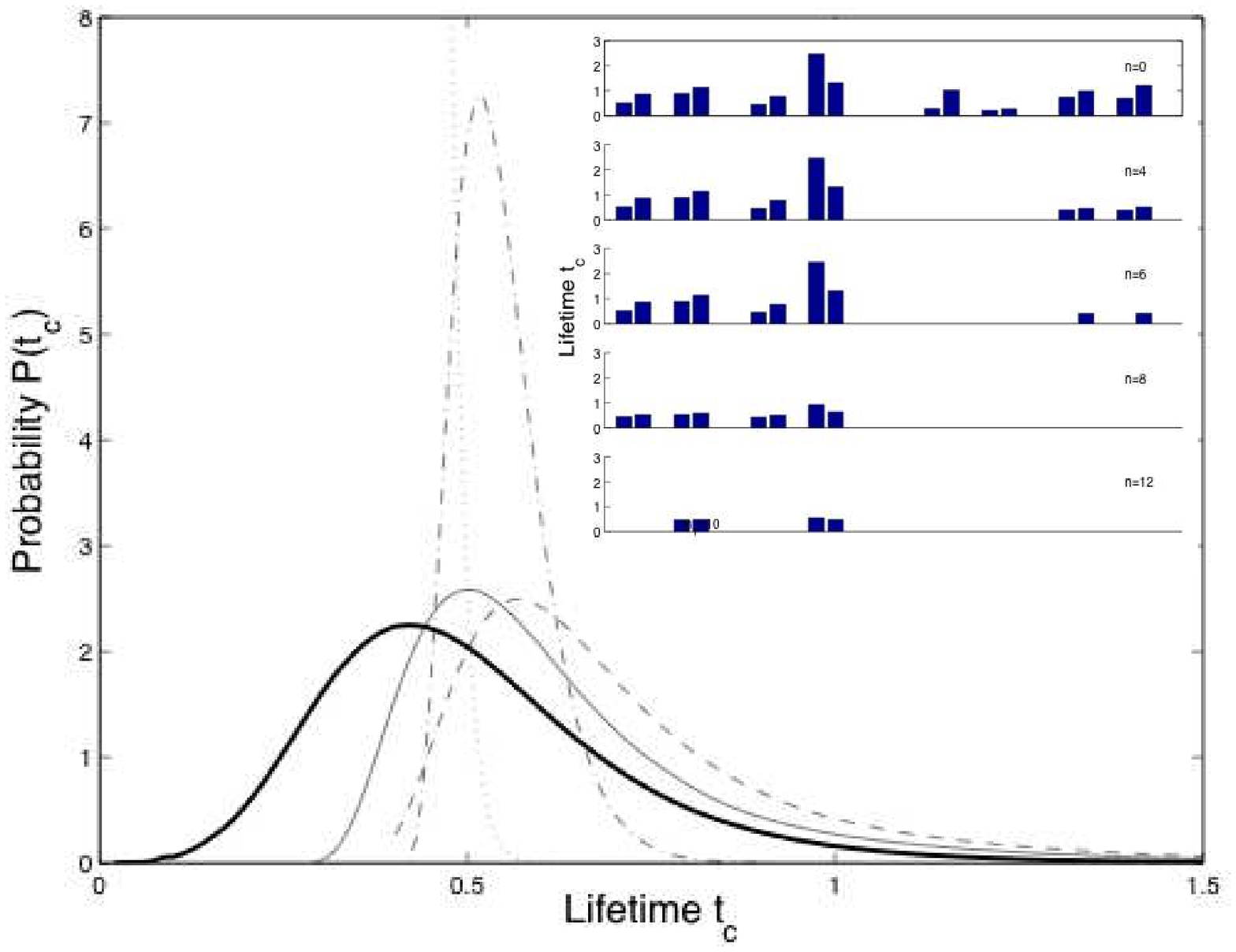}
\caption{\protect\label{Fig1}
PDFs of the lifetime $t_c$
shown at different levels of damage for a system that contained 
initially $16$ elements, just after the last element broke. The
different curves correspond to increasing numbers $n$ of broken elements:
$n=0$ (fat solid line), $n=4$ (thin solid line),
$n=6$ (dashed line), $n=8$ (dash-dotted line),
and $n=12$ (dotted line). Inset: Evolution of the
corresponding lifetimes (shown
as the bar heights) of the 16 elements with the height representing 
their lifetimes.
}
\end{figure}

We use these formulas to obtain Figure \ref{Fig1} which shows
the PDFs of the lifetime of the total system for
a different number $n$ of broken elements.
As damage is revealed, the width of the distribution decreases
which means that the
uncertainty about when the system will fail decreases. At the same time, the
most likely value of the lifetime of the system first increases up to 
$n=6$ broken elements,
after which the damage of the system is so important that a global rupture
is imminent and the most likely value of $t_c$ decreases.

\begin{figure}
\includegraphics[width=14cm]{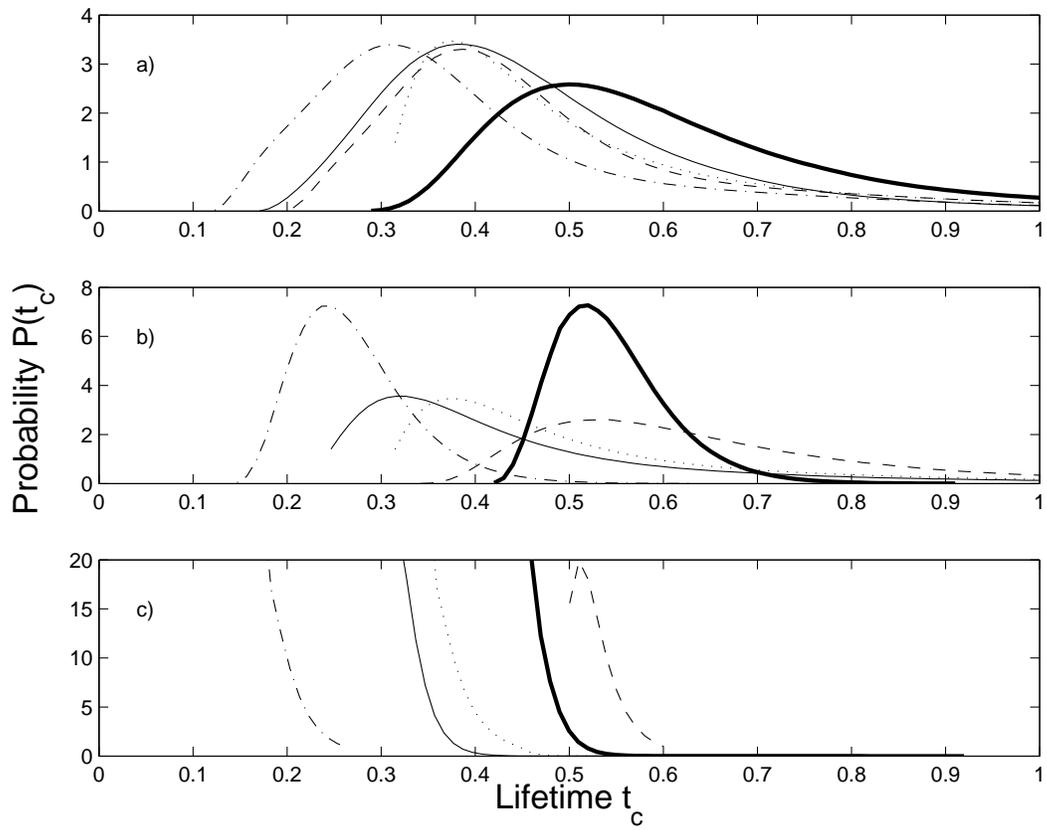}
\caption{\protect\label{Fig2}
PDFs of five different systems of $16$ elements with
different realizations of the initial lifetimes of the individual 
elements. a) $n=4$
elements broken, b) $n=8$ elements broken and c) $n=12$ elements broken.
}
\end{figure}

Figure  \ref{Fig2} illustrates the
concept of the sensitivity of the evolution of the PDF of failure times
on the initial randomness (analogous to ``chaos'' in spinglasses \cite{chaos})
and documents
two different ways by which the ``trajectories'' of two PDFs can diverge:
i) the modes (most probable value) move
apart as a function of time;
ii) the width also exhibits sensitive dependence on the quenched
randomness. Consider e.g. the PDFs represented by the
continuous and the dashed line. Their modes were slightly
different for $n=4$ broken elements but then moved closer for $n=8$
broken elements. While comparable for $n=4$, their widths have 
evolved very differently
after $n=8$ elements have failed. This illustrates the
dependence upon which sub-levels of the hierarchy which have been broken.

This prediction scheme based on incorporating iteratively the information
on the unknown pre-existing characteristic of the systems which are revealed
by the growing damage does not require a priori a complete knowledge of the
dynamics and opens the road to a suite of approximations for real 
systems involving
increasing degrees of model sophistications used in the implementation
which should be tested systematically. We expect the concept
of {\em multidimensional} dependence on
initial conditions to remain a robust
feature of the prediction of time-to-failure in many
systems, that is, there are several measures of the sensitivity to
initial conditions in the divergence of the trajectories of the
PDFs of failure times.

{}

\clearpage

\end{document}